\documentclass[12pt,preprint]{aastex}

\begin{document}

\title{The Rate of Type Ia Supernovae at High Redshift}

\shorttitle{Cosmological SNIa Rates}

\author{Brian J. Barris\altaffilmark{1} and John L. Tonry\altaffilmark{1}}

\altaffiltext{1}{Institute for Astronomy (IfA), University of Hawaii,
2680 Woodlawn Drive, Honolulu, HI 96822; barris@ifa.hawaii.edu,
jt@ifa.hawaii.edu}

\begin{abstract}
\label{batmloop-abstract}

We derive the rates of Type Ia supernovae (SNIa) over a wide range
of redshifts using a complete sample from the IfA Deep Survey.  This
sample of more than 100 SNIa is the largest set ever collected from
a single survey, and therefore uniquely powerful for a detailed
supernova rate (SNR) calculation.  Measurements of the SNR as a
function of cosmological time offer a glimpse into the relationship
between the star formation rate (SFR) and Type Ia SNR, and may
provide evidence for the progenitor pathway.  We observe a
progressively increasing Type Ia SNR between redshifts
$z\approx0.3-0.8$.  The Type Ia SNR measurements are consistent with a
short time delay ($t \approx 1$ Gyr) with respect to the SFR,
indicating a fairly prompt evolution of SNIa progenitor systems.  We
derive a best-fit value of SFR/SNR $\approx 580 \ h_{70}^{-2}$
M$_{\mathord\odot}$/SNIa for the conversion factor between star
formation and SNIa rates, as determined for a delay time of
$t\approx1$ Gyr between the SFR and the Type Ia SNR.  More complete
measurements of the Type Ia SNR at $z>1$ are necessary to conclusively
determine the SFR--SNR relationship and constrain SNIa evolutionary
pathways.

\end{abstract}


\section{Introduction}

Studies of the various types of supernovae (SN) at high redshift can
place important constraints on such processes as the star formation
rate (SFR) as a function of redshift as well as progenitor models for
the various types of SN.  Unfortunately, surveys for SN at
cosmological distances are typically concerned with finding as many
Type Ia SN (SNIa) as possible given the limitations imposed by the
difficulty in obtaining observing time with wide-field cameras and
spectrographs on large telescopes, which make it difficult to control
for systematic biases that may prevent the calculation of accurate
rates.  The standard procedure for SN surveys is to observe a patch of
sky on two nights separated by a few weeks, to allow for SN to
explode and approach maximum brightness (see, for example, Perlmutter
et al. 1995, Schmidt et al. 1998).  The images are subtracted and
searched, with objects determined to be the most likely SNIa
candidates observed with a sufficiently powerful spectrograph to
measure redshift and positively identify SN features.  The result is a
collection of SN likely to be biased from the actual distribution due
to a number of effects.  Near the sensitivity limits of any survey,
Malmquist-type biases will come into play.  This may affect SN samples
in unexpected ways if the luminosity function changes with $z$, so
that it is sampled differently at different redshifts.  The need to
select spectroscopic targets is also an important limitation on SN
surveys, because it requires that the likelihood of obtaining a useful
spectrum of the SN be a strong consideration in sample selection.
Candidates that are located at or near the centers of galaxies are
often rejected for multiple reasons.  First, they are considered
likely to be due to AGN activity, rather than a SN (particularly for
conventional surveys, where there is typically no previously
established baseline for an object's photometric behavior).  Second,
at the core of a galaxy a SN is often difficult to separate from the
background light.  These factors tend to make it unlikely that SN
candidates near the centers of host galaxies will be observed
spectroscopically.  Since it is only possible to observe a small
fraction of the discovered candidates spectroscopically, prudence
requires selecting those candidates most likely to yield positive and
unambiguous identification as a SNIa.  Again it should be noted that
these concerns become dominant at $z>0.5$, and surveys at much lower
$z$ are not expected to be as susceptible.

Despite these and other limitations in past surveys which potentially
cause the discovered samples to be significantly biased from the
actual distribution, there have been numerous attempts to estimate the
Type Ia supernova rate (SNR) over a range of redshifts.  At low
redshifts, measurements have been given by Hardin et al. 2000
($z\approx0.15$), Madgwick et al. 2003 ($z\approx0.1$), Blanc et
al. 2004 ($z\approx0.13$), and Cappellaro et al. 1999 ($z\approx0.01$).
Pain et al. (1996) was the first measurement of the rate of SNIa at
large cosmological distances, using a sample of only 3 SN at a mean
redshift of $\langle z \rangle = 0.38$ from a search area of 1.7
sq. deg.  This was extended by Pain et al. (2002), with a sample of 38
SN at a mean redshift of $\langle z \rangle = 0.55$ in an area of
$\sim$12 sq. deg.  Tonry et al. (2003) used a sample of 8 SN at a
mean redshift of $\langle z \rangle = 0.46$ from a search area of 2.4
sq. deg.  As noted by Tonry et al., the derived rates for all three of
these studies are in general agreement with a constant Type Ia SNR to
$z \approx 0.5$, indicating that observational biases may not be a
significant problem (or that the samples are all affected by the same
biases, despite coming from independent searches).  However, the
recent results of Blanc et al. (2004) hint that the SNR evolves
between $z\approx0.1-0.5$.

Generally, SN surveys are carried out with the strategy described
above, which can be termed the ``single-template'' method.  A major
limitation to these surveys is their inability to obtain light-curves
for the majority of the discovered SN, due to the difficulties in
obtaining telescope time for both spectroscopic and photometric
follow-up observations.  As discussed above, this problem becomes
particularly difficult at $z>0.5$ where the resources necessary for an
accurate spectroscopic confirmation become so great.  An alternative
approach is the ``continuous search'' strategy, which uses wide-field
imagers to repeatedly observe a region of the sky.  This allows for
more efficient discovery of SN and automatically provides full
light-curve coverage for all variable objects in the survey fields.
The first SN survey to be carried out in this manner was the IfA Deep
Survey (Barris et al. 2004), which observed an area of 2.5 square
degrees approximately every other week to a 5-$\sigma$ point source
depth of $m \approx 25$ in $RIZ$ during late 2001 and early 2002,
using Suprime-Cam (Miyazaki et al. 1998) on the Subaru 8.2-m
telescope, and the 12K camera (Cuillandre et al. 1999) on the
Canada-France-Hawaii 3.6-m telescope (CFHT).  The area was divided
into 5 fields of $\approx0.5$ sq. deg. well separated in R.A. to allow
for continuous coverage over most of the fall and winter from Mauna
Kea.  The conventional SN search component carried out while the
survey was ongoing discovered two dozen SNIa, including 15 at
$z>0.7$, doubling the then-existing published sample at these high
redshifts.  Also discovered were over 100 additional SN candidates
which were not confirmed as SNIa, most of which were not observed
spectroscopically at all (see IAU Circulars referenced in Barris et
al. 2004).

Dahlen et al. (2004) have recently measured rates based on SN found
during a SN search component of the ACS GOODS survey (Riess 2002),
which included a continuous search similar to the IfA Deep Survey.
They observed a total of 25 SNIa out to very high redshifts
($z\approx1.6$) in an area of $\sim$300 sq. arcmin, allowing them to
estimate rates over a wide range of redshifts, using bins centered at
$\langle z \rangle = 0.4$, 0.8, 1.2, 1.6.  The large number of SN
discovered allows an investigation into the evolution of the Type Ia
SNR, impossible for previous surveys which were forced to bin objects
together over a wide range of redshifts.  They report a significant
increase in the Type Ia SNR from $z=0.4$ to $z=0.8$, in contrast with
other recent measurements indicating a fairly flat SNR (e.g., Tonry et
al. 2003).  The small area of the survey means that the uncertainties
are fairly large at low-$z$ where their observed volume is small, as
well as at the high-$z$ end where incompleteness may become severe.
Using this indication of an evolution in the Type Ia SNR, in
combination with SFR measurements, Strolger et al. (2004) argue that a
long time-delay ($t\approx3-4$ Gyr) is indicated by these
observations.  With measurements of the Type Ia SNR based on the IfA
Deep Survey we are able to further investigate this question, and the
implications for SNIa progenitors.

\subsection{Type Ia Supernova Progenitor Models}
\label{rate-progenitors}

The specific progenitors of SNIa are still unknown, though there is
a general consensus that they are the result of the explosion of C-O
white dwarfs (see Hillebrandt \& Niemeyer 2000; as well as Nomoto et
al. 2000 and other work from the same volume, for several recent
reviews).  This is a worrying situation to be in due to the
profound implications for modern astrophysics indicated by the
evidence for an accelerating universe based on SNIa observations
(e.g. Barris et al. 2004, Tonry et al. 2003, Knop et al. 2003, Riess
et al. 1998, Perlmutter et al. 1999).

It is now agreed that there must be multiple pathways to a SNIa
explosion in order to account for the diversity of their observed
light-curve and spectral properties (as demonstrated by Li et
al. 2001).  Discovery of a SNIa within a known progenitor system
would be a major breakthrough, but given the low rate of occurrence
and the faintness of the assumed progenitors, such a felicitous event
is unlikely in the near future (though see Ruiz-Lapuente et al. 2004
for a possible identification of a binary progenitor system for Tycho
Brahe's 1572 supernova).  Reliance on theoretical models (or
plausibility arguments, such as those made by Livio \& Riess 2003) is
not desirable, both because of significant model uncertainties, such as
specific information regarding the composition of the progenitor and
accreted material and the details of the explosion, and because of a
fundamental uncertainty about whether a model that might produce
accurate predictions of SNIa observables may actually correspond to
the pathway occurring in nature.

One significant difference between the DD and SD models is their
expected evolutionary timescales.  Stars that produce C-O WDs have
main-sequence masses of $3-8$ M$_{\mathord\odot}$ (Nomoto et
al. 1994), and therefore have a short lifetime ($t < 0.6$ Gyr).  A DD
system can therefore form within a few tens to hundreds of millions of
years after the initial period of star formation that produced the
progenitor stars.  The size of the orbit of the two C-O WDs then
determines the length of time until they coalesce.  One common model
for the formation of a DD system involves a common-envelope phase (see
Yungelson \& Livio 2000 and references therein), which will result in
quite close orbits and short times until the SN explosion.  The
estimated length of time between the initial star formation ($not$ the
formation of the DD system) and the explosion as a SNIa peaks at
roughly a few hundred million years (Yungelson \& Livio 2000,
Ruiz-Lapuente et al. 2000).  For a SD system, the companion star can
be fairly low mass, with a correspondingly long main-sequence
lifetime.  The time until SN explosion is therefore determined by the
evolutionary timescale of the companion star, so that a distribution
of $t > 1$ Gyr is expected (Yungelson \& Livio 2000, Ruiz-Lapuente et
al. 2000).

This difference between the two classes of models indicates that a
detailed study of the SFR and Type Ia SNR could assist in determining
the preferred SNIa pathway (see Ruiz-Lapuente et al. 1995, Madau,
della Valle \& Panagia 1998, Gal-Yam \& Maoz 2004, and the reviews
given above).  If the Type Ia SNR closely follows the SFR, then the
progenitor evolution must be short, and the DD pathway is likely to be
preferred (based on current understanding of the two models).  If the
Type Ia SNR appears to be significantly delayed with respect to the
SFR, then the evolutionary timescale is long, indicating that the SD
model is more likely to reflect reality.  The measurement of accurate
SNIa rates, despite the inherent difficulties discussed above, may
therefore be a powerful tool for resolving our current lack of
understanding of the details of these explosions.

\subsection{Metallicity Measurements as Probes of SNR}

There have been indications supporting a short time delay between
progenitor formation and SNIa explosion from a number of metallicity
studies.  The different classifications of SN play different roles in
the chemical enrichment of galaxies and the intracluster medium (ICM)
due to the differences in explosion products.  The bulk of oxygen and
a fraction of iron is produced in SN II, while SNIa produce the
majority of iron (see Matteucci \& Greggio 1986, and references
therein), with the details dependent upon the IMF, which determines
the fraction of stars that end up as each of the various SN types.
Observed metallicity ratios such as [O/Fe] $vs$ [Fe/H] can therefore
aid in constraining the relationship between Type Ia SNR and Type II
SNR (which is expected to very closely the SFR, due to the very short
lifetimes of massive stars that end up at Type II SN).  Pettini et
al. (1999) found that metallicity ratios reveal that damped Ly$\alpha$
systems show signs of enrichment from SNIa, indicating that
substantial numbers have exploded at redshifts higher than $z \approx
1.5$, which points to a short delay-time.  Matteucci \& Recchi (2001),
using [O/Fe] and [Fe/H] measurements from the solar neighborhood,
derive a timescale of $t\sim1$ Gyr at which SNIa become the dominant
source of Fe production, though they note that this value need not be
universal.  Scannapieco \& Bildsten (astro-ph/0507456), also using
[O/Fe] ratios, predict a SNR dominated by a prompt component with a
time delay of $t\sim0.7$ Gyr.

Kobayashi et al. (1998) developed a model in which the Type Ia SNR is
strongly dependent upon the Fe abundance found within the progenitors,
and predict a decline between $z \sim 1-2$, which would match that
predicted by the long delay favored by Strolger et al. (2004), whereas
a short delay results in a Type Ia SNR that does not peak until $z>2$.
In further work, Kobayashi, Tsujimoto, \& Nomoto (2000) predict a
global Type Ia SNR which is quite different from the long-delay model
(and simple short delays as well), exhibiting a second peak at $z \sim
3$ due to evolution of the local environment and dominant progenitor
pathway throughout cosmological history.  Measurements of the SNR at
such high redshifts are not yet possible.

In this paper we detail the calculations of SNIa rates from the IfA
Deep Survey.  In Section ~\ref{chapter-lcoe} we describe the process
by which we have re-analyzed the IfA Deep Survey data to provide a
first look at the very faint, variable universe, and a comprehensive
catalog of SN candidates in the survey fields, necessary for an
accurate calculation of SN rates.  In Section ~\ref{sample-select} we
apply several tests designed to weed out potential non-Type Ia
contaminants to produce a sample of 98 SN which we are confident is a
complete sample of SNIa with a minimal number of interlopers.  In
Section ~\ref{measuring-rates} we describe the process of determining
the underlying SN rate based on the number of SN that were detected.
Section ~\ref{rate-calculation} presents the calculated rates and
discusses some of the complications.  Section ~\ref{sec-models}
details implications of comparisons with star formation rates, and
Section ~\ref{rate-conclude} gives our conclusions.

\section{Light Curves of Everything}
\label{chapter-lcoe}

As mentioned above, the IfA Deep Survey, with its repeated use of
wide-field imagers on large telescopes, had several advantages over
traditional surveys carried out with the ``single-template'' strategy.
First, the exclusive use of wide-field cameras meant that each survey
field could be searched for new SN after every new observation.  This
allows a very large sample to be accumulated (since we are in a sense
undertaking numerous conventional surveys consecutively) and therefore
enables the calculation of SNR over a range of statistically
reasonably-sampled redshift bins.  Second, complete light-curves were
obtained for all discovered SN rather than for a select few.  This is
crucial since the large number of discovered objects precludes
obtaining spectroscopic observations for the entire sample, so
light-curve shape information must be used in many instances for both
SN identification and redshift determination.  Third, after the
conclusion of the survey light-curves spanning the time baseline of
several months can be constructed for $all$ objects in the fields,
allowing a complete study of variable sources in the survey area.  We
call this ability ``Light Curves of Everything,'' or LCOE.  Similar
work has recently been presented by Becker et al. (2004), from Deep
Lens Survey observations that also made repeat visits to selected
fields over extended periods of time.

The most obvious application of LCOE for our purposes is to augment
the SN search component of the IfA Deep Survey.  Our extensive set of
SN detection software is readily extendable to the task of producing
and searching an arbitrarily large set of difference images for
variable sources.  Upon identification of SN candidates, our
light-curve analysis procedure may be readily performed.  The primary
difficulty lies in the intermediate step of determining which
photometrically variable objects are indeed SN, from the overwhelming
number of other types of objects.  During the course of the IfA Deep
Survey, and in other SN surveys, this task was accomplished by
searching every subtraction by eye.  This is required because
automated object-detection algorithms typically have a high rate of
false-positives, as well as missing many real objects of interest, and
attempts to suppress the former usually lead to a larger number of the
latter.  However, manual inspection during a complete LCOE re-analysis
is utterly impractical for several reasons.  First, we wish to search
$all$ pairs of images, rather than a single difference image,
increasing the amount of data to be searched by multiple orders of
magnitude.  In addition, we would like to control systematics in a
meaningful way, and so wish to remove human subjectivity from the
process as much as possible.

\subsection{Detection of Photometrically Variable Objects}
\label{lcoe-detection}

The basic procedure for detecting variable objects is the same as that
used in conventional SN surveys (see, e.g., Schmidt et al. 1998, Tonry
et al. 2003, Barris et al. 2004) for discovering SN---for a given
pair of images, PSF matching is performed, the images are placed onto
a common flux scale, and the subtractions performed.  Object detection
software is then run on the difference image in order to detect
variable sources, which are more easily identified in such an image.
During the SN survey, difference images from a pair of observations
are then inspected by human searchers in order to find the best
candidates, which are then observed spectroscopically.

We wish to perform a more comprehensive search for variable objects,
and so this process was repeated in a much more general fashion.  For
every pair of images, the subtraction and search steps were carried
out, and a catalog of variable sources detected by our search software
was constructed.  We perform all $N*(N-1)$ possible subtractions,
rather than the $N*(N-1)/2$ number of unique combinations, because we
wish to find both rising and declining objects in all epochs, and our
detection software does not trigger on ``holes'' from negative
variations.  The catalogs from every subtraction are then merged into
a master catalog of all variable objects in each field.

This process was repeated for each of the three survey filters
($RIZ$).  The union of the catalogs from each filter was taken in
order to produce a master list of all objects which exhibited
variability in any of the three filters.  Using this master list,
photometry in each filter was rerun for every object to construct a
final list of complete multi-band photometry for all variable sources.
Even though a given object might have been detected in only a single
filter, flux measurements were taken at the source position for every
subtraction in every filter.  Light-curves were then constructed using
the NN2 method described by Barris (2004) and Barris et
al. 2005 (astro-ph/0507584).  This method performs all $N*(N-1)/2$ distinct
subtractions possible from $N$ observations, and uses the resulting
matrix of flux differences to construct variable object light-curves,
in contrast with the ``single template'' method which defines a single
observation as the template and subtracts it from the other $N-1$
observations (i.e., forming a single column of the NN2 flux difference
matrix).  The ``single template'' method is highly sensitive to errors
caused by imperfections in the template image, while the NN2 method is
not as dependent on any single image being of high quality.

The result of the pipeline procedure was a catalog containing a total
of more than 160,000 potential variable sources.  A full analysis of
these objects, in order to classify them based on light-curve and
image properties and comprehensively study the faint, variable
universe, is a prodigious task worthy of more attention than can be
dedicated here.  We will concentrate on using this sample to further
our studies of SNIa.

\subsection{Testing for Completeness with Simulated Supernovae}
\label{fakers}

During the LCOE search process, we inserted simulated supernovae into
the survey images in order to objectively measure the sensitivity of
our objection-detection software.  We inserted more than 3000
simulated supernovae into each field throughout the entire course of
the survey, mimicking SNIa light curves.

Each field was divided into 48 patches of size 2k x 2k pixels, and in
each patch 100 ``fake SNIa'' were inserted.  These ``fakers'' each
had a light-curve roughly approximating that of a SNIa, increasing
and decreasing in brightness with peak magnitude ranging from very
bright ($m=21$) to extremely faint ($m=26$), timed to peak at
different epochs during the survey and with positions stepped by
subpixel amounts to check for PSF-matching systematic errors.  Once
inserted, the ``fakers'' were treated with the same detection,
photometry, and cataloging procedures as true targets.

This procedure allows us to probe in detail the performance of our
object-detection pipeline.  Our use of the simulated SN in
combination with the LCOE analysis provides a test of both our object
detection software and our tests for extracting likely SN candidates
from the variable source catalog (see Section ~\ref{lcoe-tests}).
Barris (2004) includes detailed quantitative results for the
determination of completion efficiency.  Most important is the
completeness limit, which we define as the magnitude at which the
number of detected simulated supernovae drops to one-half the total
inserted number.  This value is typically approximately 24.1 for $I$
band, 24.3 for $R$ band, and 23.4 for $Z$ band.

\subsection{Extracting Likely Type Ia Supernovae}
\label{lcoe-tests}

From the overwhelming number of discovered objects, we wish to
efficiently identify likely SNIa, since individually examining the
images or light-curves of all of these objects is impossible.  In
order to facilitate the analysis of detected objects, a total of
twelve parameters were calculated for each object with an eye towards
their use to isolate SN from other types of variable objects.  These
parameters include image properties such as the FWHM of the variable
source; the magnitude of the variable source and any underlying
source, such as the host galaxy of a SN candidate or AGN; and the
distance between the location of the variable source and the ``host''
object.  Other parameters describe additional light-curve properties,
and include the overall light-curve RMS; the ratio between the RMS and
the background flux level; the RMS based on observations from each
individual instrument; the RMS when excluding the single most
discrepant point; the average RMS between pairs of subsequent
observations; the difference between the maximum and minimum observed
flux values; the number of times the light-curve crosses the midpoint
between the maximum and minimum observed flux values; the length of
time between these crossings; and the number of observations in which
the object was detected.

It is not immediately obvious what values we should expect a SN
light-curve to have for our calculated image and light-curve
parameters.  However, we possess a few dozen excellent examples of SN
Ia in the 23 presented by Barris et al. (2004).  We can use these
objects as a ``training set'' and define areas of parameter space
where they, and presumably SNIa in general, are likely to lie.  For
many of the calculated parameters, the 23 IfA Deep Survey SNIa
generally lie in a more compact region of multi-parameter space than
the complete set of variable objects, indicating that it is possible
to construct tests to isolate SNIa with these parameters.  

After these tests were applied, a total sample of 727 potential SN
candidates remained, a small enough number that no further pruning was
considered necessary.  At this point the insertion of human judgment
into the process was unavoidable, as these objects had to be inspected
by eye to categorize their light-curves.  Several broad categories
were obvious upon inspection of the sample:

1. SN-like: Objects which appear consistent with a SN light-curve.
These may have a galaxy ``host'' or no visible host at all.  In the
ideal case, a rise and fall are observed, over a plausible time scale.
See Barris et al. (2004) for 23 examples of this class of object.  For
many objects, however, the full light-curve evolution does not fall
within the time-frame of the survey, so that only a rise or fall is
observed.  Such objects are often not suitable for our analysis since
their identification as SN is not compelling (see below).

\begin{figure}
\plotone{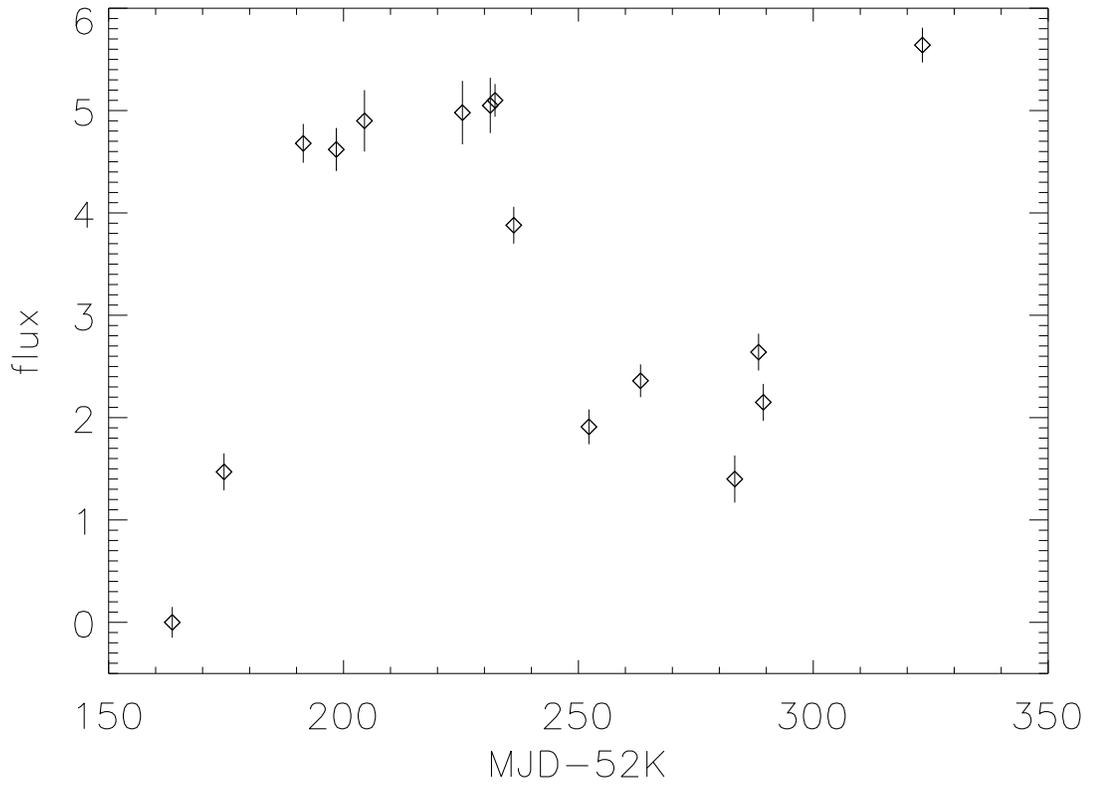}
\caption{Light-curve of example prototype object classified as AGN-like.}
\label{agnlc}
\end{figure}

\begin{figure}
\epsscale{0.5}
\plotone{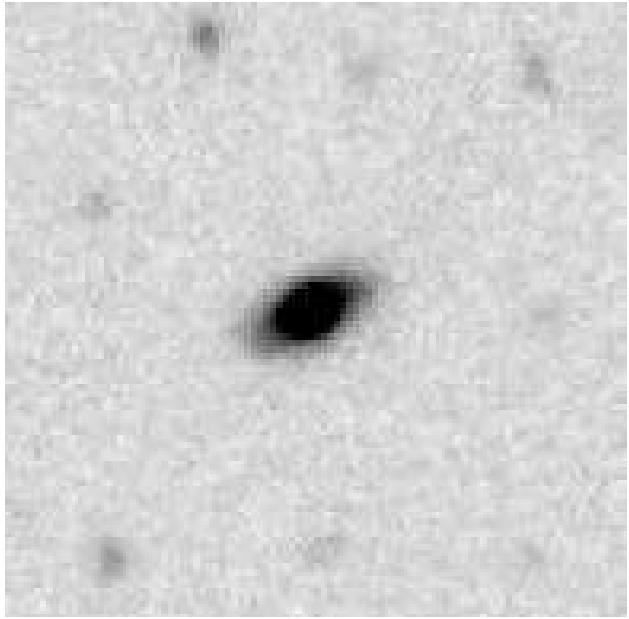}
\caption{Image of prototype object classified as AGN-like (light curve given in 
Figure ~\ref{agnlc}).}
\label{agnpic}
\end{figure}

2. AGN-like: Objects which do not have a SN-like light curve, and
appear to have a galaxy ``host'' rather than a point-like ``host.''
They exhibit a wide range of light-curve properties (e.g., varying
between two plateaus of brightness; exhibiting variability timescales
not appropriate for a SN of the observed brightness; showing no
obvious pattern to the light-curve but varying at levels beyond what
we believe is our photometric accuracy).  See Figures ~\ref{agnlc} and
~\ref{agnpic} for a light-curve and image of an example object
classified as ``AGN-like.''

\begin{figure}
\plotone{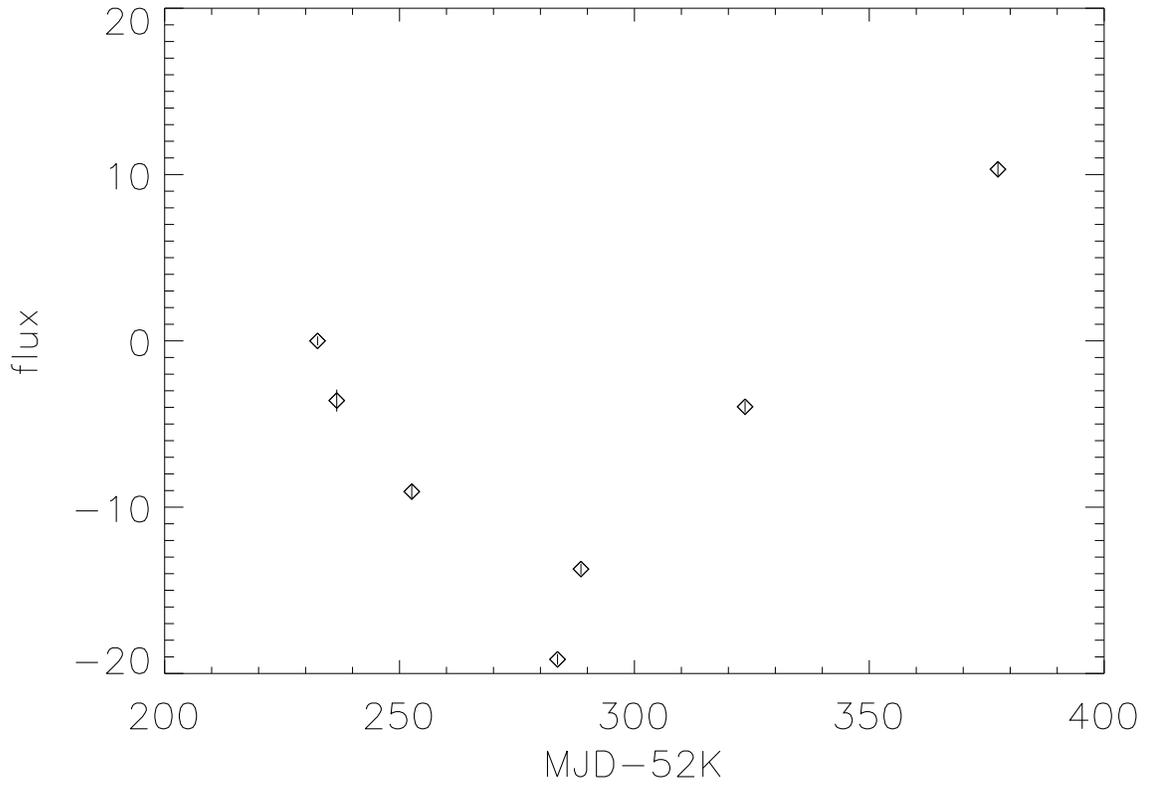}
\caption{Light-curve of example prototype object classified as Variable-star-like.}
\label{vslc}
\end{figure}

\begin{figure}
\epsscale{0.5}
\plotone{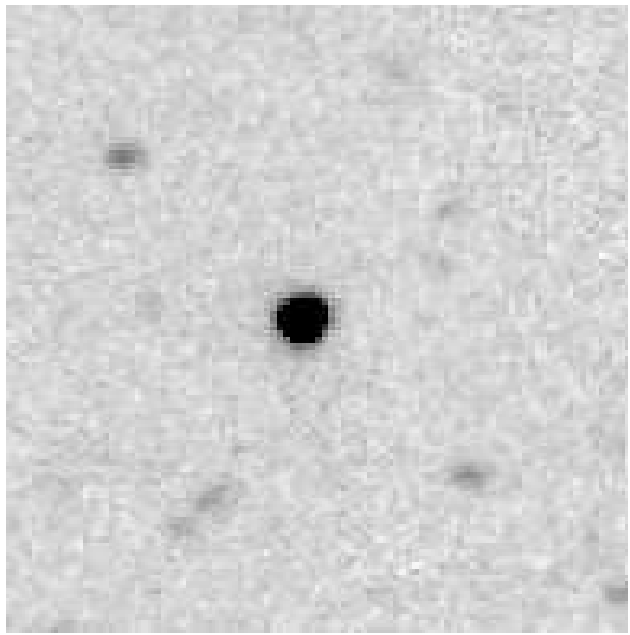}
\caption{Image of prototype object classified as Variable-star-like (light curve given in 
Figure ~\ref{vslc}).}
\label{vspic}
\end{figure}

3. Variable-star-like: Objects with light-curves similar to those
described for AGN-like candidates, but with a stellar appearance
(which does not necessarily preclude them from being AGN).  A number
exhibit some apparent periodicity in their light curve.  See Figures
~\ref{vslc} and ~\ref{vspic} for a light-curve and image of an example
object classified as ``Variable-star-like.''

4. Technical issues: Variability arising from various problems with
the images and/or subtractions rather than actual astrophysical
phenomena of interest (diffraction spikes; moving objects, such as
satellites and solar system objects, that are incompletely removed by
cosmic-ray rejection software; objects which fall too near an edge, so
that different epochs include them fully, partially, or not-at-all on
the mosaic; etc).

A total of 172 objects were classified as SN-like, further separated
into three different subgroups.  First, there are the 23 SNIa
discovered and spectroscopically confirmed during the IfA Deep survey.
Second, 43 objects that were discovered during the survey but for
which no spectroscopic information was obtained (see Barris et
al. 2004, as well as the IAU Circulars referenced therein).  These are
objects for which our confidence in their identification as SNIa is
high based on their multi-color light-curves, but which were not
spectroscopically confirmed due to the limitation in spectroscopic
observing time relative to imaging.  The LCOE investigation
essentially ``re-discovered'' these 43 objects, as well as the 23 IfA
Deep Survey SNIa.  Finally, 106 objects which are newly discovered
from the LCOE analysis.  Because our tests recovered $all$ of the 23
known SNIa (by construction), as well as numerous additional SN
that were also previously known, we are confident that our search
criteria are reliable and that our sample of SN from the survey area
is complete to a magnitude of $m_I\approx24.1$.

\section{Sample Selection}
\label{sample-select}

We describe above the process by which we have re-analyzed the IfA
Deep Survey observations to extract a complete sample of SNIa
candidates from the survey fields.  This set of 172 SN candidates
includes a number of objects with incomplete light-curves, whose full
rise and decline is not sufficiently sampled by the survey
observations.  We consider objects insufficiently covered if they
occur so early that there is no clear indication that the maximum was
observed, or so late that there are not multiple detections on either
side of the peak.  This constraint is carefully taken into account
below in our calculation of the control time of the survey.  Culling
such objects leaves 133 candidates, divided among the three groups
mentioned above in numbers of 23/40/70.  Barris (2004) lists
positional and other information for these 133 SN candidates,
including redshift-independent distances measured with the method
described by Barris \& Tonry (2004).  This procedure calculates
luminosity distances by marginalizing over redshift, which is
necessary since we do not posses redshift information for the large
majority of our SN.  We perform light-curve fitting over a range of
redshifts using the Bayesian Adapted Template Match (BATM) method (a
non-parametric SNIa light-curve fitting procedure first described by
Tonry et al. 2003 and Barris et al. 2004, and in more detail by Barris
2004), and marginalize over $z$, expanding on conventional distance
measurement techniques that often marginalize over parameters such as
extinction and time-of-explosion while requiring $z$ to be an input
parameter.

The BATM method for calculating luminosity distances for SNIa is
based upon an idealized set of representative SNIa light-curves that
are photometrically and spectroscopically well-sampled in time, and
for which accurate distances are known.  With such a
spectrophotometric template set, predicted light-curves could be
produced to compare to observations.  However, data of this quality
are extremely uncommon at present, so we use a set of light-curves
which have excellent temporal coverage over a range of wavelengths and
span a wide range of luminosity, and a large set of observed spectra.
For a given redshift, the SEDs are shifted and warped so that they
match the observed photometry of each template light-curve.  BATM
treats the ``template'' and ``unknown'' in a fundamentally different
manner from previous methods for measuring luminosity distances to SN
Ia (Phillips 1993; Riess et al. 1996, 1998; Perlmutter et al. 1997).
The SEDs and light-curves are shifted to the redshift of the SN to be
measured, so that redshift effects are $introduced$ to the template
set rather than $removed$ from the observational data.  The idea is to
compare the observed SN to what we would expect the template to look
like at a given redshift, as opposed to comparing the template to what
the SN would look like were it at the redshift of the template.  BATM
compares the unknown SN to a set of 20 template light-curves (see
Barris 2004), with the final answer for distance calculated by
combining the results from all of the templates based on probabilistic
weights calculated from a $\chi^{2}$ test.

Barris \& Tonry (2004) demonstrated that for a large sample of 60
Hubble-flow SNIa, $z$-independent distances have approximately the
same scatter relative to the Hubble diagram as those using
conventional distance methods.  Since we wish to determine rates as a
function of redshift, we must convert the distances measured by Barris
(2004) into corresponding values of redshift.  To do so one must adopt
a cosmology.  We have chosen to use cosmological density parameters of
($\Omega_M, \Omega_\Lambda$) = (0.3, 0.7).  In addition to being the
currently favored model, this allows the most straightforward
comparison with recent authors who have measured rates using similar
values.

While the light-curve coverage provided by the extended time baseline
of the observations makes us confident that these objects are likely
SN, for the majority we do not have spectroscopic confirmation of
their SN classification.  We know that our sample selection criteria
described above are effective at locating SNIa, since they were
based upon our known sample of 23 such objects, but also expect that
some fraction of the 133 may not be SNIa.  We only took spectra of 2
SN II during the course of the IfA Deep Survey, and neither of these
objects is found within our sample.  This is encouraging because it
indicates that many SN II did not pass our tests and are not in the
sample (which we attribute to the very different shapes of many SN II
light-curves; see Filippenko 1997), though it is likely that some are.
We wish to construct further tests to determine which objects are
likely to be non-SNIa, so that we may remove them from our sample
before calculating rates.

There are a few tests that we can perform that are analogous to those
that were used by Barris et al. (2004) to confirm whether candidates
were indeed SNIa.  The first is a goodness-of-fit test.  As a part
of the redshift-independent BATM procedure used to measure distances,
light-curves are fit at a wide range of redshifts, and at each we have
a measure of the goodness-of-fit ($\chi^{2}/N$).  This value varies
over the redshift range as certain combinations of redshifted spectral
and light-curve templates match the observed light-curve better than
others---for example, a faint, broad light-curve will be a much better
fit at high-$z$ than low-$z$, and hence have a smaller $\chi^{2}$.  If
a light-curve has large $\chi^{2}$ values over the $entire$ redshift
range, it indicates that the observations are not a good match to a SN
Ia at $any$ redshift, and the object may not belong in our sample.

The second test concerns the colors of the SN.  If an object has
colors consistent with a SNIa, it is further evidence that it is
likely to be one (see, for instance, Barris et al. 2004 and Riess et
al. 2004).  However, any light-curve fitting procedure can deal with
unusual colors by adjusting the extinction $A_{V}$ of the fit.
Extreme values of $A_{V}$ may be a warning sign that the observed
colors are not a good match to a SNIa, though care must be taken to
consider the fact that some legitimate SN may in fact actually be
heavily extinguished.  Objects that are too blue, and therefore are
fit with a negative extinction, are difficult to reconcile with any
legitimate physical process, and may be rejected without too much
concern that they are actually SNIa.  One concern may be that if a
SN in the BATM training set has not been properly de-reddenned,
comparison with an unknown SN may reveal it to be overly blue,
implying a negative extinction.  To prevent such confusion, every
attempt has been made to use only SNIa with low extinction values as
templates.

In much the same way that we used the 23 known SNIa as a ``training
set'' in Section ~\ref{lcoe-tests} to define regions of
multi-parameter space where such objects lie, we can use them as a
guide to define ``normal'' behaviors of $\chi^{2}$ and $A_{V}$ for SN
Ia, and reject only candidates that lie well outside of such values.
With these guidelines, we have imposed a cutoff rejecting SN which
never have values of $\chi^{2}/N<4$ over the redshift range of
$z=0.2-1.2$.  This eliminates 13 objects.  We reject a further 6
objects because they have $\chi^{2}/N>4$ at redshifts near the values
corresponding to their calculated luminosity distance in an
($\Omega_M, \Omega_\Lambda$) = (0.3, 0.7) universe, and achieve
$\chi^{2}/N<4$ only for significantly different redshifts.  For
example, if a SN whose distance corresponds to $z=0.8$ has
$\chi^{2}/N=5$ at that redshift, but $\chi^{2}/N=2$ at $z=0.2$, it is
an indication of something strange with the observed light-curve, and
it may not actually be a SNIa.  In examining color, we reject
candidates with fit values of $A_{V}>1$ when evaluated at the redshift
corresponding to their distance.  This eliminates 6 objects.  Finally,
we reject 5 candidates with values of $A_{V}<-0.05$ when evaluated at
this redshift, leaving 103 SN.  This cutoff value was chosen to
eliminate the large number of objects seen in larger samples with very
negative extinction values.

An additional test is to look more closely at the light-curve fits for
each SN, and examine the template light-curves that match the
observations the best.  While the final answer for BATM distance is
calculated by combining the results from all of the templates, the
template that matches best can provide insight into the general
light-curve shape.  Our selection of light-curve templates reflects
the fact that SNIa are a very homogeneous set of objects, but we
have included two extremely subluminous Type Ia (SN 1999by and SN
1991bg) to account for the nontrivial number of such faint objects in
the luminosity function.  However, these SN are so faint ($\sim 1.5$
magnitudes fainter than ``normal'' Type Ia---see Filippenko et
al. 1992, Vink{\' o} et al. 2001) that we do not expect to find them
above very low redshifts.  If a $z\approx0.5$ or higher SN is best fit
with one of these faint templates, it is likely a sign that something
is amiss, since it would be evidence for a fast-declining, bright SN
Ia, a type of object that is not believed to exist.  When we
investigate the templates that are the best match to every SN in each
of the four highest redshift bins, there are 4 matched to SN 1999by or
SN 1991bg in the $z=0.55$ bin, none in the $z=0.45$ and $z=0.65$ bins,
and 1 in the $z=0.75$ bin.  Rejecting these 5 objects yields a sample
of 98 SN, which we will consider our complete sample of SNIa for
the purpose of deriving the observed rates at high-$z$.

\begin{figure}
\plotone{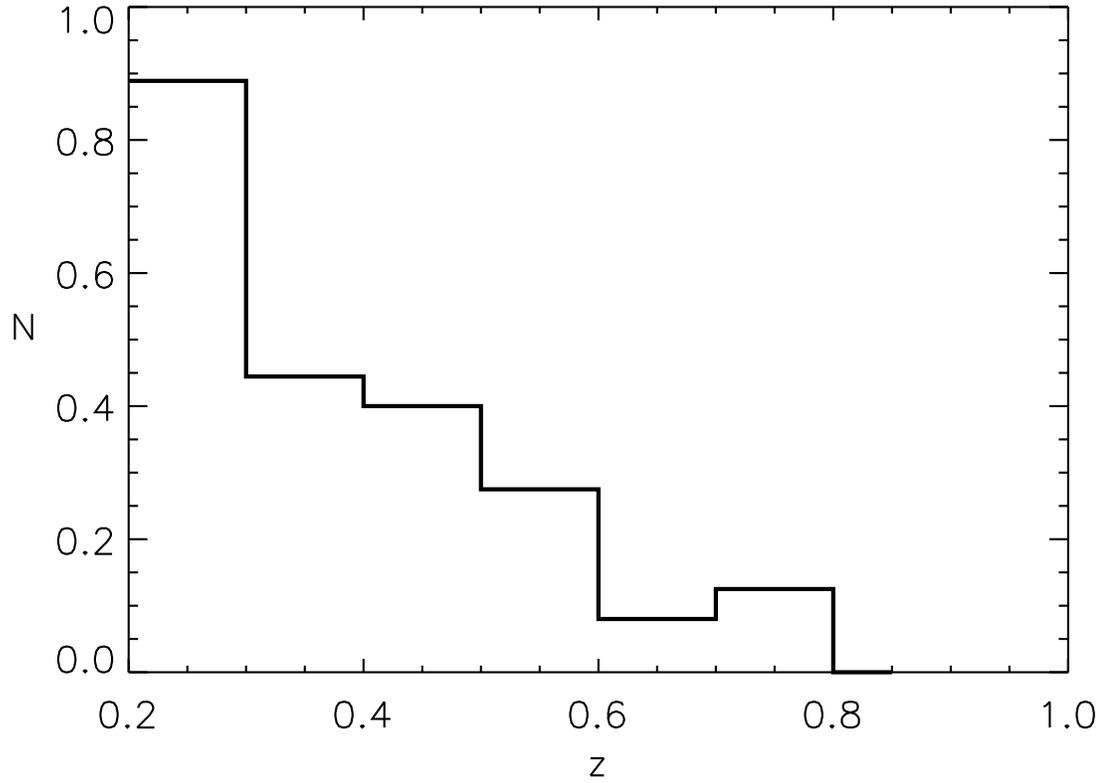}
\caption{Histogram of the fraction of SN from each bin that were
rejected during the culling process from 133 to 98 SN.  Note the
decline in the rejected fraction to high-$z$, which we attribute to
the fainter luminosity function of the most likely potential
contaminants (primarily CC SN).}
\label{cullhist}
\end{figure}

Figure ~\ref{cullhist} plots the fraction of objects that were removed
from each redshift bin during the initial weeding-out process to cull
the sample from 133 to 98 candidates.  The bins at low-$z$ have a
higher fraction of rejected objects than the bins at high-$z$.  We
believe this is due to an actual increased level of contamination at
the bright end of our initial sample of 133 objects.  Likely
contaminants include non-Type Ia SN (see below for further
investigations into this possibility), variable stars, and AGN.  We
attribute the observed drop in the rejected fraction to the fainter
luminosity function of the most likely contaminants (Type II and Ib/c
SN being typically much fainter than SNIa; see Richardson et
al. 2002) causing them to fail to be found by our search at high-$z$,
while those at low-$z$ that do make it into our sample are fit poorly
and are rejected by our tests (hence the high rejection rate).

\section{Measuring Rates}
\label{measuring-rates}

In addition to the sample of SN to use, there are several other
pieces of information that must be determined before a rate
calculation is possible.  These include parameters pertinent to the
IfA Deep Survey such as the area covered, sensitivity, and time
coverage; properties of SNIa such as luminosity function, light
curve shape; and expected host-galaxy extinction values.  With this
information, we will be able to relate the number of SN observed to
the total number of SN that actually occurred, as a function of
redshift, and hence measure SNR($z$).

First we must determine the sensitivity limits of the survey, to know
how deep we probed in magnitude.  Rather than simply examining the
signal-to-noise ($S/N$), seeing, and other image parameters, and
assuming that we were sensitive to $S/N=10$ or some such value, we
carried out a procedure that allows an objective assessment of how
well we recovered, as a function of magnitude, simulated SN inserted
into the survey images (see Section ~\ref{fakers} above).  We will use
the magnitudes corresponding to the 50\% completeness limit as a
measurement of the sensitivity of the LCOE search procedure that was
used to construct our sample of SN.  This is equal to $m=24.1-24.2$
for the five survey fields.  Note that this is $not$ the detection
limit for a typical single-night observation, which is approximately
$m_I=25.2$ for 5-$\sigma$ point source sensitivity.

Next we must calculate the effective area covered by the survey.  This
is complicated by the fact that we observed with both CFHT+12K and
Subaru+SuprimeCam, with different spatial coverage for the two
instruments (see Barris et al. 2004).  Each of the five survey fields
was observed with two overlapping SuprimeCam pointings ($\approx0.5$
sq. deg.) but only a single 12K FOV ($\approx0.38$ sq. deg.).  SN
located in the ``core'' region imaged by both Subaru and CFHT have a
different length of time during which they are detectable compared to
those in the ``border'' regions, which are only imaged by Subaru.  In
effect we have in each field two separate surveys that must be
considered in predicting the expected number of observable SN.
Adding up all these regions, and allowing for a loss of 10\% due to
bad pixels, edge effects, and other image defects (a value that we use
based on prior experience, and consider to be an upper limit), we
calculate a total area of 2.3 sq. deg.  The effects of the different
time coverage of the various fields and sub-fields are discussed in
more detail below.

We must consider values for the SNIa luminosity function in order to
determine the magnitudes and numbers of SN that we can expect to
detect.  We adopt the luminosity function for the peak brightness of
SNIa given by Li et al. (2001), consisting of three distinct
subclassifications, each individually modeled by a Gaussian
distribution.  These three subgroups are ``normal'' SNIa with mean
absolute magnitude $M_B=-19.3$ mag (for $h=0.70$), a dispersion of
$\sigma=0.45$~mag, and accounting for 64\% of events; ``bright'', SN
1991T-like objects with $M_B=-19.6$ mag and $\sigma=0.30$~mag and
accounting for 20\% of events; and ``faint'', SN 1991bg-like objects
with $M_B=-17.8$ mag and $\sigma=0.50$~mag, accounting for 16\% of
events.  During the rate calculation, the normalization of the
luminosity function is matched to the observed number of SN to
determine the underlying SNR.

Next we require an estimate of the control time covered by the survey.
A SN of a given combination of brightness, redshift, and explosion
time will be detectable during the period of time when it is brighter
than our detection limit.  It is the sum of these times that we are
sensitive to that is of interest, $not$ the total amount of survey
exposure time.  Consider a standard single-template SN survey.
Subtracting the template epoch from the second (``search'') epoch
yields a difference image suitable for discovering SN.  A SN that
exploded over a range of days between the observations will be
detectable, so that if it is visible for a week, then $that$ is the
time sensitivity (control time) of the survey (since, reversing the
perspective, it means that we are sensitive to SN that explode during
a period of one week) rather than the two nights on which we observed
or the few weeks time between observations.

The calculation of the control time covered by the survey is
complicated by the fact that SN have different values for intrinsic
brightness and may occur at various redshifts.  In order to calculate
the number of SN that we expect to have been able to detect, we work
in the parameter space of ($z$, $m_{max,I}$).  To calculate the SNIa
density in this space, we first calculate how many SNIa occur at
each point.  We use the value of $m_{max,I}$ from SN 1995D with an SED
from SN 1994S to convert to the corresponding value in ($z$,
$m_{max,B}$) space, which allows us to use the adopted luminosity
function to determine the expected number of such objects.  We
multiply in a volume factor of $dV/dz/d\Omega$, working with the flat,
$\Lambda$-dominated universe with ($\Omega_M, \Omega_\Lambda$) = (0.3,
0.7), and also multiply by a time-dilation factor of $(1+z)^{-1}$.
This is necessary because we are interested in the predicted rate for
$observing$ SNIa (what we measure) given a rest-frame rate (which we
desire).  Next we convolve the distribution with a model for the
effects of dust extinction approximating the results of Hatano,
Branch, \& Deaton (1998).  In their model 25\% of hosts are ``bulge
systems'' with the probability for host $A_B$ described by $f(A_B)
\propto 0.02\,\delta(A_B) + 10^{-1.25-A_B}$, and 75\% are ``disk
systems'' with $f(A_B) \propto 0.02\,\delta(A_B) + e^{-3.77-A_B^2}$.
We use redshifted SEDs to determine how the host-galaxy extinction
$A_B$ determines the observed $I$-band extinction.  The assumption
that the host galaxy extinction does not vary with redshift is
certainly incorrect at some level.  However, it would require
substantial additional color information to attempt to satisfactorily
disentangle redshift and host galaxy extinction, and so we have made
the simplifying assumption of no variation.

Given the density of SNIa within ($z$, $m_{max,I}$) space, we take
into account that for different values of these parameters a SN is
visible to an observer for different periods of time.  At a given
redshift, a bright SN will obviously be above the detection threshold
for a longer period of time than a faint SN due to the differences in
peak brightness, and the difference is enhanced by the fact that
brighter SN have broader light-curves (e.g., Phillips 1993).
Redshift affects the length of time a SN is visible by broadening the
light-curve, by dimming it through increased distance, and by changing
the portion of the spectrum which lies within the observed filter.
For each point in ($z$, $m_{max,I}$) space we must consider how long a
SN with those values may be observed.  Note that even though the LCOE
search procedure was run on all three survey filters ($RIZ$), it is
valid to consider that the search was effectively performed only in
$I$-band, since the design of the survey was such that the S/N in the
$I$-band images was much higher than for $R$ and $Z$, so SNIa should
be more detectable in this filter than in the other filters at all
redshifts.

To calculate the control time from the SNIa densities in ($z$,
$m_{max,I}$) parameter space, we adopt two well-observed SNIa as
light-curve templates---SN 1995D, a bright SN with $\Delta m_{15} =
0.99$ and MLCS $\Delta = -0.42$; and SN 1999by, a faint SN with
$\Delta m_{15} = 1.93$, MLCS $\Delta = +1.48$.  We use a similar
procedure to that used within the BATM method to place these objects
in ($z$, $m_{max,I}$) space.  At a given redshift, light-curves for
SN corresponding to values of $m_{max,I}$ between the two templates
are determined by a flux interpolation between them.  Light-curves
brighter than SN 1995D are calculated by adding a constant positive
magnitude offset to SN 1995D to match the peak brightness, and those
fainter than SN 1999by are determined by adding a negative offset to
it.  With these light-curves, we can determine how long a SN at any
value of ($z$, $m_{max,I}$) is above the threshold magnitude limit of
our survey.  Traditional single-template surveys require one to
consider that a SN may have exploded before the first observation,
meaning that the flux in the subtraction image is the difference
between the fluxes in the two epochs, rather than the flux present in
the second observation.  The long time baseline covered by the IfA
Deep Survey and our use of the NN2 procedure (Barris et al. 2005,
astro-ph/0507584) mean that we always have an observation where the SN
is not present, so we always have a difference image that is sensitive
to the actual SN flux.

\section{Supernova Rate Calculation}
\label{rate-calculation}

We are now equipped with the necessary information to calculate the
rates of SNIa that are consistent with the observed yield from the
IfA Deep Survey.  We will simply assume that the luminosity function
is constant with redshift, i.e. that the relative fractions of bright,
normal, and faint SNIa do not change from the values given above, so
that the normalization (in terms of number of SNIa per volume per
year) is the only free parameter for determining the underlying rate
through comparison with the observed sample.

\begin{figure}
\includegraphics[scale=.6,angle=-90]{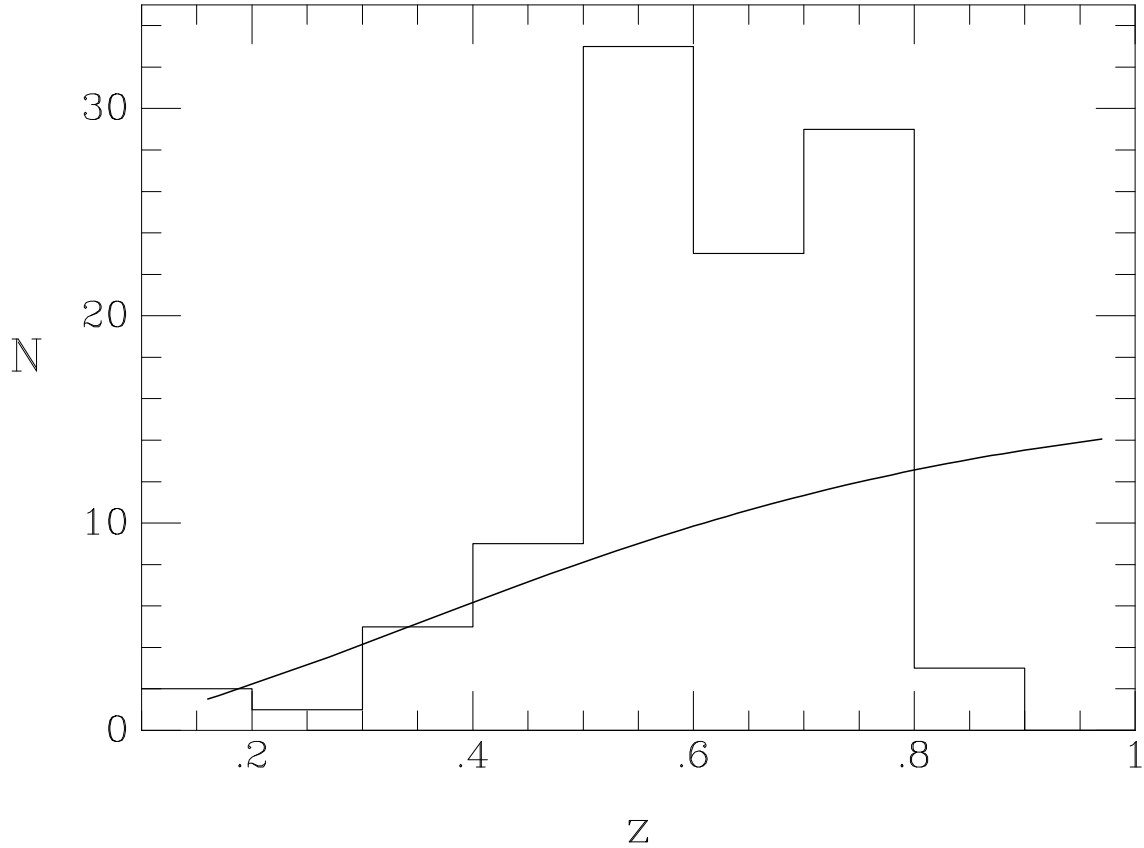}
\caption{Histogram of the number of discovered SN as a function of
redshift.  Also shown is the envelope expected to describe the
histogram (with arbitrary normalization) were it merely a reflection
of the changing volume and redshift being sampled.  The steep rise of
the histogram with increasing redshift indicates that the numbers of
discovered objects are indeed a sign of an actual increase in the
SNR.}
\label{hhzss_z}
\end{figure}

In Figure ~\ref{hhzss_z} we plot a histogram of the number of SN
within bins of a width of $z=0.1$, calculated from the distances given
in Barris (2004) for a ($\Omega_M, \Omega_\Lambda$) = (0.3, 0.7)
cosmology.  We also show the curve giving the trend of $dV/dz/(1+z)$,
which could, with proper normalization, describe the expected envelope
of the histogram if the rise in number of SN was simply a reflection
of the increased volume and redshift being sampled at different
points.  The IfA Deep Survey sample discovered is $not$ well described
by the increasing volume and time/redshift envelope, indicating that
the $z>0.5$ bins in Figure ~\ref{hhzss_z} are truly a sign of enhanced
SNR.  In order to quantify the effect, we now calculate the actual
rate corresponding to these histograms.  We will truncate our
calculation with the $z=0.75$ bin, beyond which sample incompleteness
apparently becomes quite severe.

In Table ~\ref{table:rates} we present the SNR calculations.  We have
used the regular redshift bins of width $z=0.1$ shown in Figure
~\ref{hhzss_z}, and give the number of objects within each bin for the
initial sample of 133 as well as the final sample of 98 SN.  We also
include the number of SN that we expect the IfA Deep Survey to have
been able to detect, given the information and assumptions from the
previous section, if the rate were equal to a constant value of
10$^{-4}$ SNIa Mpc$^{-3}$ yr$^{-1}$ $h_{70}^3$, enabling us to
determine the rate corresponding to the numbers that were actually
observed.  For uncertainties we adopt Poisson N$^{1/2}$ errors for
each bin.  Values for both rates and uncertainties are obtained by
scaling by the expected number of SN in each redshift bin.

\begin{figure}
\includegraphics[scale=.7,angle=-90]{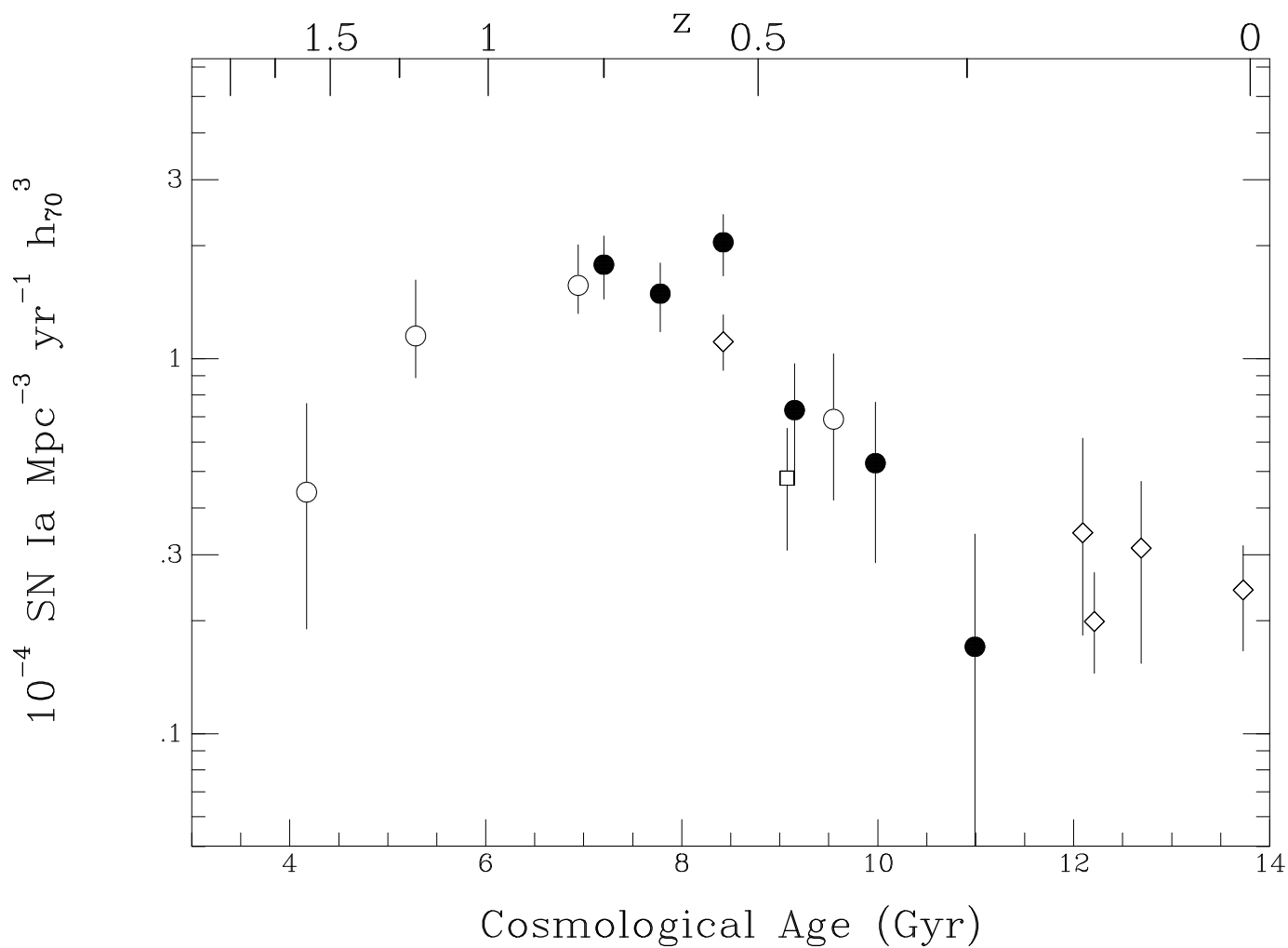}
\caption{Rates from the IfA Deep Survey (large filled circles) as a
function of cosmological time, compared with previously published
results: Tonry et al. 2003 (open square); Dahlen et al. 2004 (open
circles); open diamonds represent, with increasing redshift,
Cappellaro et al. (1999), Madgwick et al. (2003), Blanc et al. (2000),
Hardin et al. (2000), and Pain et al. (2002).  Note the large
discrepancy between the IfA Deep Survey measurement and that of Pain
et al. at $z=0.55$.  This may indicate sample contamination of this
redshift bin.}
\label{ratescomparemod}
\end{figure}

In Figure ~\ref{ratescomparemod} we plot these results, as well as
recent values from the literature for comparison.  The large number of
SNIa discovered during the IfA Deep Survey enables us to examine the
Type Ia SNR as a function of redshift to an unprecedented level of
detail.  The rates calculated at $z=0.35$ and $z=0.45$ are in good
agreement with the recent calculations presented in the literature,
indicating that our sample selection and calculation procedure appear
to be working properly.  Similarly, the derived value at $z=0.75$
agrees well with that at $z=0.80$ given by Dahlen et al. (2004),
confirming their observation of a significantly higher rate at
$z\approx0.8$ compared to $z\approx0.45$.  Our observed point at
$z=0.65$ also appears to describe a fairly gradual increase in the SNR
as a function of redshift, peaking at perhaps $z\approx0.8-1.0$, with
the points between $z=0.65-0.80$ consistent with a flattening in the
rate over this redshift range.

The point at $z=0.55$ is somewhat of an outlier to such a picture.
Rather than displaying the general trend of gradually increasing SNR
to higher redshifts, the rate at $z=0.55$ is sharply higher than that
at $z=0.45$, and higher than the rate measured for $any$ higher
redshift bin.  It is also substantially higher than the value reported
by Pain et al. (2002) at the same redshift, with which it is
discrepant at $>3\sigma$ level.  We have investigated several
systematic effects that could potentially cause this point to be
biased to such a high value.  First, the redshift-independent BATM
method that we have used may tend to incorrectly measure distances for
our sample SN, scattering objects into this bin.  Among objections to
this theory are the good agreement the higher-$z$ and lower-$z$ bins
exhibit with previously published rate measurements and the large size
of the discrepancy.  If there were a large number of SNIa being
improperly placed into the $z=0.55$ bin, it seems likely that moving
them into their proper bins would create a discrepancy in these other
bins where none currently exists.  We have also checked the derived
redshifts for the 23 known SNIa from Barris et al. (2004), and no
spike at $z=0.55$ is observed, a reassuring sign that the $z$-free
BATM distances are not systematically overpopulating the bin with SN
Ia from other bins.  Another possibility is that there are non-SNIa
in the sample which tend to be fit at distances corresponding to
$z\approx0.5$.  The most obvious candidates for interlopers are CC SN
(either Type Ib/c or Type II), though AGN and flaring variable stars
could also perhaps mimic SNIa light-curves in certain cases.  SN II
have a very broad (and uncertain) luminosity function (Richardson et
al. 2002), but are roughly two magnitudes fainter than SNIa.  We do
not believe that SN II are plausible contaminants, however, because
their light-curve shapes are quite different from those of SNIa,
which will cause most SN II to be rejected by our goodness-of-fit
tests, or even be rejected during the LCOE analysis (as for the 2 SN
II that we know of that are not included even in our sample of 133).
SN Ib/c have light-curve shapes similar to Type Ia, and also have a
much fainter luminosity function, so it is possible that they could
fool our fitting procedure.  To test this, we have measured $z$-free
distances for SN 1994I, a typical SN Ic with significant extinction.
The light-curves do tend to be fit at higher redshifts, but for this
particular SN are not placed in the $z=0.55$ bin.  The most important
revelation is that the derived fit colors are extremely blue.  Such
objects would not be able to contaminate our sample, because we would
reject it based on such unusual colors.  We therefore do not have an
obvious source for contamination in the $z=0.55$ bin, and conclude
that the rate measurement at this redshift is accurate.  We also
expect that the analysis that we have performed for this bin gives a
general sense of the potential systematic effects on any individual
bin, which we also do not expect to be severe.

There is strong justification for the expectation that the systematic errors  are small, certainly in comparison with the statistical errors.  The primary sources arise from the possibility that the assumptions that were made in Section ~\ref{measuring-rates} are incorrect.  The most obvious potential discrepancies include the evolution of the assumed dust extinction model, which is tied to an assumed distribution of SN Ia in different types of host galaxy, and the uniformity of the spectrophotometric templates that we have used, which affects both the maximum brightness and the calculations K-corrections.  However, the well-established properties of SN Ia are such that effects such as these should be small.  Their high degree of heterogeneity implies that our templates should be sufficient for an accurate rate calculation.  As has been shown by Dahlen et al. (2004) and others, the value of dust extinction typically needs to be increased by a factor of several before its effects are comparable with statistical errors for similar SN rate calculations.

The primary concern that we have about the accuracy of our rate measurements is for incompleteness at high redshifts.  It is clear that our redshift bins at $z>0.75$ are highly incomplete.  Attempting to fully quantify the potential systematic errors that may afflict the lower-$z$ bins, e.g. through detailed Monte Carlo simulations of the effects of slight variations in dust extinction, will have much less impact on the conclusions that may be drawn from the measured SN rates than will the recognition that the highest-$z$ bins are incomplete to such an extent that the measurements should be disregarded.  As we show in the next section, we believe that the conclusions drawn by Strolger et al. (2004) may be entirely driven by measurements that are highly incomplete rather than reflecting actual evolution in the Type Ia SNR.

\section{Comparison with Star Formation Models}
\label{sec-models}

The SNR measurements shown in Figure ~\ref{ratescomparemod} are in
general described by a gradual increase in the Type Ia SNR with higher
redshift, with signs of flattening by $z\approx0.8$, perhaps even as
early as $z\approx0.6$.  The calculations from the IfA Deep Survey
agree well with the recent results from Tonry et al. (2003) and Dahlen
et al. (2004), and confirm the indication from the latter that there
is a substantial increase in the SNR between $z\approx0.45$ and
$z\approx0.8$.  The size of our sample enables us to trace the history
of the SNR in more detail than previously possible.  We would like to
compare the observed SNR to measurements of the SFR in order to
determine whether insight can be gleaned into potential progenitor
pathways.

We can first quantify the relationship between SFR and the Type Ia
SNR.  We have chosen to use the SFR values of Steidel et al. 1999 in
order to facilitate a comparison with the recent results presented by
Strolger et al. (2004).  As shown by the recent compilation of Hopkins
et al. (2004), measurements of cosmological SFR at
different wavelengths are so widely discrepant that it is not at all
clear how well the Steidel et al. values reflect the actual underlying
star formation.  However, Strolger et al. (2004) recently used the
Steidel et al. values to draw profound conclusions about the
relationship between SFR and SNR, so we have chosen to do so also in
order to facilitate a comparison with their results.  We use all SNR
measurements at $z<1$ from the surveys mentioned previously.  We do
not use the higher $z$ values from Dahlen et al. (2004) for two
reasons--first, neither the IfA Deep Survey nor any other survey has
measured SNR at such high redshifts, so there is no confirmation of
the apparent turnover in the SNR indicated by these values.  Second,
the results of Strolger et al. (2004) indicating a long delay time
between SFR and SNR are strongly driven by the measurements at high
$z$, so much so that the addition of our measurements at $z<0.8$ will
not change their conclusions.  We wish to test whether there is any
evidence for the Strolger et al. conclusions other than the two
highest $z$ redshift bins measured by Dahlen et al.

Using these two collections of SNR and SFR values, we fit over a range
of delay times for the SFR/SNR ratio that best matches the
observations, using a least-squares analysis.  We interpolate the SFR
measurements in order to compare with the SNR, rather than choosing a
parameterization such as that defined by Giavalisco et al. (2004).  We
have chosen to do so because it is not clear that this
parameterization accurately reflects the published SFR measurements.
For example, the SFR measurements of Steidel et al. (1999) are
constant between $z=0.75-1.76$ ($t\approx4-7$ Gyr), yet the model
indicates an increase of nearly a factor of 3 over this range.
Considering the large observational uncertainties (and the cosmic
variance which may afflict the SFR measurements), it is not clear that
the model of Giavalisco et al. (2004) should be preferred over other
models, including one produced directly from the observations, for
comparing with the Type Ia SNR.

\begin{figure}
\includegraphics[scale=.7,angle=-90]{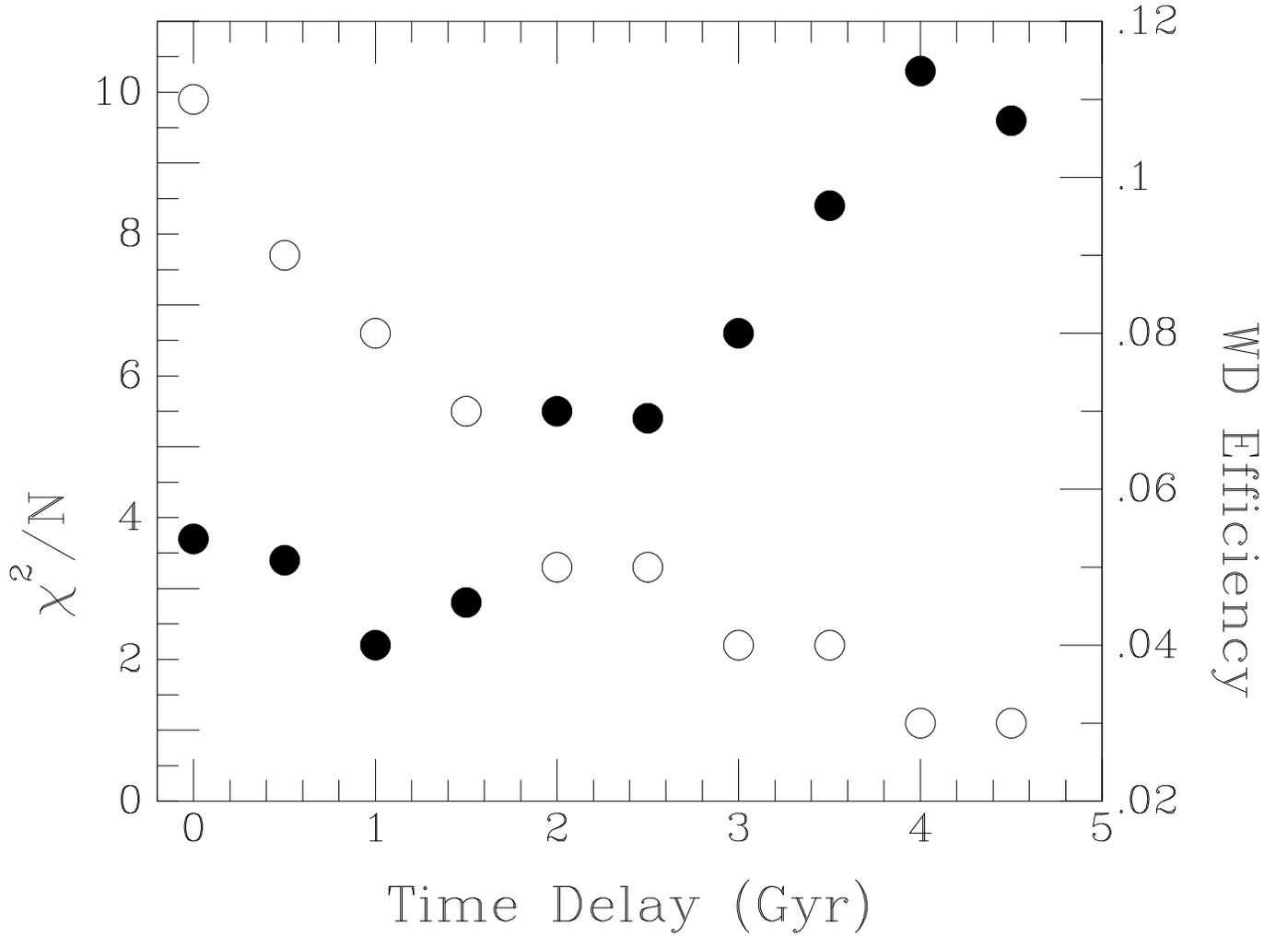}
\caption{Trends in $\chi^{2}/N$ (filled circles) and C-O WD explosion
efficiency (open circles) as a function of delay-time.  Note the
best-fit values for a match between SFR and SNR measurements occurs at
a time delay of $t\approx1$ Gyr.  Longer time delays produce
substantially poorer fits.}
\label{table2fig}
\end{figure}

\begin{figure}
\epsscale{.6}
\plotone{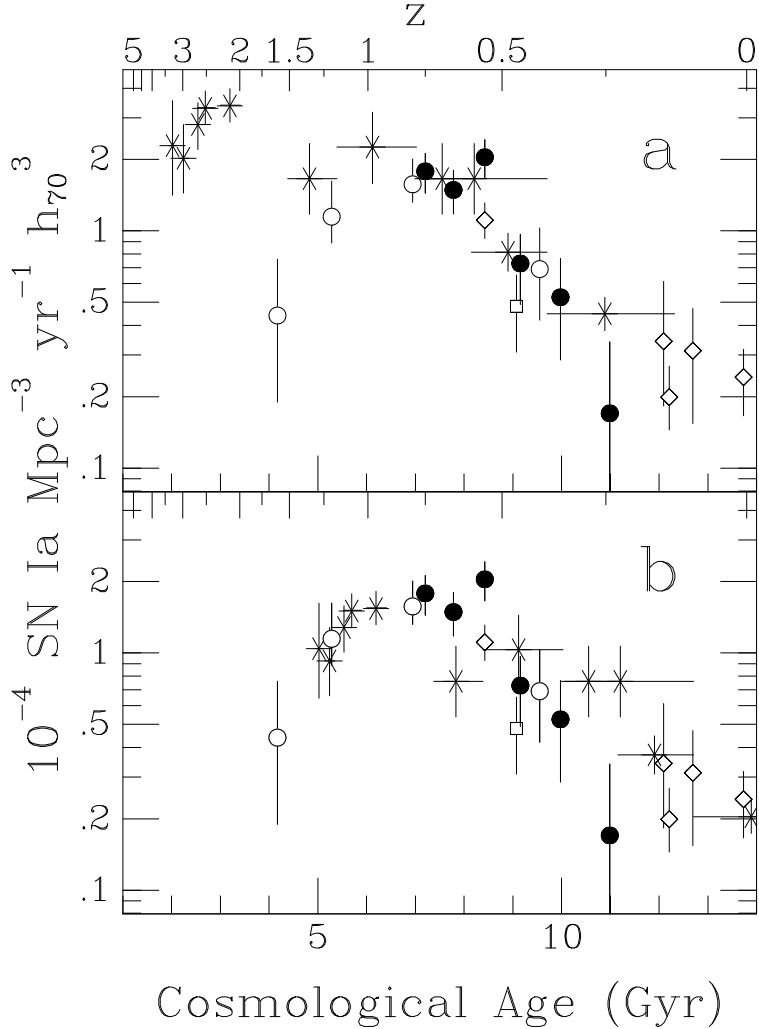}
\caption{As in Figure ~\ref{ratescomparemod}, Type Ia SNR calculations
from the IfA Deep Survey (large filled circles), and previously
published rates as given in Figure ~\ref{ratescomparemod}.  Also
included in this figure are SFR measurements from Steidel et al. 1999
(asterisks, scaled to overlie the SNR values).  a) A simple linear
shift of $t=1$ Gyr is introduced to the SFR observations as indicated
by Table ~\ref{table:sfr22}.  Note the common plateau feature between
$t\approx5-8$ Gyr, as well as the subsequent decline seen in both SFR
and SNR measurements.  b) A time-delay of $t=4$ Gyr is introduced to
the SFR observations as suggested by Strolger et al. (2004).  This
match is produced by aligning the apparent turnover in the SNR
beginning at $z\approx1.2$ with SFR measurements at $z\approx3-6$.
However, note the poor match between the actual SFR and SNR
observations between $t\approx8-12$ Gyr ($z\approx0.75-0.25$) in
comparison to the linear shift of $t \approx 1$ Gyr shown in a).  }
\label{twobanger}
\end{figure}

The best agreement between SFR and SNR measurements occurs at
$t\approx1$ Gyr, with a derived ratio of SFR/SNR = 580 $h_{70}^{-2}$
M$_{\mathord\odot}$/SNIa (see Table ~\ref{table:sfr22} and Figure
~\ref{table2fig}; it should be noted that the deviation of the
calculated $\chi^{2}/N$ from 1 is solely due to the $z=0.55$ bin).  In
Figure ~\ref{twobanger}a we compare the SNR and SFR measurements,
shifting the latter by $t=1$ Gyr.  This time delay reflect the
best-fit value from Table ~\ref{table:sfr22}, and Figure
~\ref{twobanger}a demonstrates that it matches features evident in
both observations---a sudden drop at $z\approx0.55$ ($t\approx8.4$
Gyr) in the SNR measurements, after previously being relatively
constant for a long period of time ($t\approx3$ Gyr or more).  While
we have discussed the possibility of sample contamination afflicting
our SNR measurement at $z\approx0.55$, it may indeed be a sign of a
short ($t\approx1$ Gyr) delayed reaction to a similar relatively
abrupt change in the star formation rate, with the repeated caveat
that such features in the SFR are even more uncertain than for the
SNR.

In Figure ~\ref{twobanger}b we plot the SFR and SNR observations, with
a long delay time of $t = 4$ Gyr.  This value is the best-fit model
given by Strolger et al. (2004) for the ACS data and previously
published SNR measurements, determined through a comparison with the
Giavalisco et al. (2004) SFR model.  It is obvious that the apparent
turnover indicated by the point at $z=1.6$ is the driving force behind
the findings by Dahlen et al. (2004) and Strolger et al. (2004) of a
long time-delay between the SFR and SNR.  An examination of Figure
~\ref{twobanger} reveals that the more detailed redshift coverage
provided by the large IfA Deep Survey sample does not agree as well
with this model as for a short delay time.  As discussed above, the
flattening in the SNR (indicated by measurements from both the IfA
Deep and ACS Higher-Z Surveys) between $z=0.5-1.2$ is well matched by
a similar plateau in the SFR $t\approx1$ Gyr earlier.  In Figure
~\ref{twobanger}b, however, the constant region in the SFR
measurements has been shifted to correspond to redshifts between
$z=0.25-0.7$, a region during which the SNR declines substantially.
This reveals the reason for the significantly worse fits with
increasing time delay values given in Table ~\ref{table:sfr22} and
shown in Figure ~\ref{table2fig}---the slope indicated by the SFR
measurements is simply too shallow in comparison with SNR values at
times $t > 2$ Gyr later (though the match with the local SNR
measurement of Cappellaro et al. 1999 is somewhat at odds with this
finding).  While the Giavalisco et al. (2004) model curve utilized by
Strolger et al. (2004) fits the data reasonably well, especially when
constrained by the $z=1.6$ ACS measurement, the match between the
actual SFR and SNR observations is poor.

\subsection{WD Progenitor Efficiency}
\label{cowd}

We can also compare the values of SFR/SNR to the expected number of
white dwarfs produced during star formation.  We adopt a Salpeter
(1955) initial mass function (IMF) with a slope of $m=-2.35$ and a
mass range of $0.1-125$ M$_{\mathord\odot}$.  As noted in Section
~\ref{rate-progenitors}, C-O WDs are produced by stars with a
main-sequence mass of $3-8$ M$_{\mathord\odot}$ (Nomoto et al. 1994).
We can therefore integrate over the IMF to calculate the number of
such white dwarfs that are produced per unit star formation:
$$N(C-O \ WD) = \int_{3 M_{\mathord\odot}}^{8 M_{\mathord\odot}}
\psi(M) dM \ / \int_{0.1 M_{\mathord\odot}}^{125 M_{\mathord\odot}} M
\psi(M) dM,$$ which results in a value of N(C-O WD)=0.021
M$_{\mathord\odot}^{-1}$.  Table ~\ref{table:sfr22} includes values
for the number of C-O white dwarfs formed by the amount of star
formation indicated by the SFR/SNR ratio.  For time delays $t=0-1$
Gyr, there are $\approx10-12$ C-O WDs for each SNIa (adopting
$h=0.7$), which indicates an efficiency of 8--10\% ($\pm$4\%) for
producing SNIa (see Figure ~\ref{table2fig}).  This value is somewhat
higher than the 5--7 \% determined by Dahlen et al. (2004), though
within the uncertainties.

The various models for SNIa explosion mechanisms discussed in Section
~\ref{rate-progenitors} can in general all accommodate the delay time
of $t\approx1$ Gyr derived here.  In fact, even the long delay time of
$t=3-4$ Gyr preferred by Strolger et al. (2004) cannot strongly rule
out most progenitor pathways, due to theoretical uncertainties.  A
substantial amount of work must be done to further investigate this
matter.  The measurements presented here should guide the work of
theorists and modelers into providing further constraints.

\subsection{Predicting the Type Ia SNR at $z>1$}

Based on the above SFR/SNR ratios, we can predict the SNR that we
expect at higher redshifts assuming that SNIa do indeed follow the
SFR after a delay of $t\approx1$ Gyr.  At $z=1.76$ $(t\approx3.8$
Gyr), the SFR is 0.09$_{-0.03}^{+0.04}$ M$_{\mathord\odot}$ Mpc$^{-3}$
yr$^{-1}$ $h_{70}$.  Shifting this point by $t=1$ Gyr would place it
at approximately $z=1.35$.  If we adopt SFR/SNR = 580$\pm$240
$h_{70}^{-2}$ M$_{\mathord\odot}$/SNIa (for a $t=1$ Gyr time delay),
we obtain a predicted Type Ia SNR of 1.5$\pm$0.8 x 10$^{-4}$ SNIa
Mpc$^{-3}$ yr$^{-1}$ $h_{70}^3$, in good agreement with the value of
1.15$_{-0.26}^{+0.47}$ x 10$^{-4}$ SNIa Mpc$^{-3}$ yr$^{-1}$
$h_{70}^3$ measured by Dahlen et al. (2004).

The next measurement of the SFR at higher redshift is at $z\approx3$
($t\approx2.2$ Gyr), which corresponds to $z\approx2$ after a shift of
$t=1$ Gyr.  This measurement of 0.18$_{-0.03}^{+0.03}$
M$_{\mathord\odot}$ Mpc$^{-3}$ yr$^{-1}$ $h_{70}$ would predict a Type
Ia SNR of 3.0$\pm$1.3 x 10$^{-4}$ SNIa Mpc$^{-3}$ yr$^{-1}$
$h_{70}^3$ using SFR/SNR = 580$\pm$240 $h_{70}^{-2}$
M$_{\mathord\odot}$/SNIa.  This strongly conflicts with the trend
suggested by the $z=1.6$ ACS SNR measurement of 0.44 x 10$^{-4}$ SNIa
Mpc$^{-3}$ yr$^{-1}$ $h_{70}^3$.  However, it agrees well with the
predicted range of 1-3.5 x 10$^{-4}$ SNIa Mpc$^{-3}$ yr$^{-1}$
$h_{70}^3$ given by Scannapieco \& Bildsten 2005 (astro-ph/0507456).

Further measurements of the SNR at $z>1$ must be performed in order to
confirm the apparent plateau and decline indicated by the ACS GOODS
Survey.  We note that since it was a continuous survey, an analysis
similar to the LCOE process described here can be carried out on the
survey observations.  This would be helpful to overcome the
difficulties inherent in constructing a complete sample under the time
constraints of SN surveys (recall the discussion in the introduction),
and ensure the accuracy of their high-$z$ SNR measurements.  Without
clear evidence for a turnover, there is little power to differentiate
between a shift in amplitude and a shift in time, so both short ($t
\lesssim 1$ Gyr) and long ($t \approx 3-4$ Gyr) time delays can be
accommodated.  Confirmation of a turnover would be a strong indication
of the need for a long ($t\approx3-4$ Gyr) time delay between SFR and
SNR (see Strolger et al. 2004).  Using the actual SFR measurements
rather than the parametrized model of Giavalisco et al. (2004)
produces a substantially better match to the SNR measurements for
short delay times.  More measurements at $z>0.8$ are necessary before
the behavior of the SNR can be determined at these redshifts.  At
present, the observations are consistent with a fairly short ($t
\approx 1$ Gyr) time delay.

\section{Conclusion}
\label{rate-conclude}

The IfA Deep Survey, with its continuous observations of large areas
of sky, provides a unique opportunity for a general exploration of the
astronomical time domain at very faint magnitude limits and over a
wide range of timescales.  We have analyzed the survey images with
software designed to detect all variable objects and extract a
complete sample of SNIa, constructing tests which efficiently detect
variable objects with SN-like light-curves.  From an initial number of
variable objects of a few tens of thousands per field, complete to
$m_I\approx24.1$, a sample of approximately 700 candidates was
selected based upon these tests, reduced to 133 SN candidates upon
visual inspection, nearly 100 of which were determined to be useful
for cosmological analysis.  The thorough LCOE analysis process we have
implemented has allowed us to collect an extremely large, complete SN
sample extending to $z\approx0.8$.

We have investigated the rates of SNIa indicated by the complete
sample obtained from the IfA Deep Survey.  The large size and broad
redshift range of our sample makes it uniquely powerful for probing
the SNR throughout cosmological history.  We find good agreement
between our calculated rates and those previously reported at
$z\approx0.45$ and $z\approx0.8$, confirming the recent findings of
Dahlen et al. (2004) that the SNR rises substantially during this
period of cosmological history.

Through a comparison with measurements of the SFR over a wide range of
redshifts, we observe no evidence for a significant delay between the
SFR and the Type Ia SNR.  The best-fit values are for a time-delay of
$t\approx1$ Gyr, in contrast to the much longer values of $t=3-4$ Gyr
suggested recently based on results from the ACS GOODS SN survey
(Dahlen et al. 2004, Strolger et al. 2004).  This short value is
supported by metallicity observations in the solar neighborhood as
well as in high-$z$ damped Ly-$\alpha$ systems.  We are unable to
conclusively determine a preferred progenitor system, since most
theoretical models can accommodate a delay of $t\approx1$ Gyr between
the SFR and Type Ia SNR.

We derive a best-fit value of SFR/SNR $\approx 580 \ h_{70}^{-2}$
M$_{\mathord\odot}$/SNIa for the conversion factor between star
formation and SN rates, as determined for a delay time of $t\approx1$
Gyr between the SFR and the Type Ia SNR.  With this value we predict a
SNR value at $z\approx1.25$ in agreement with the published results of
Dahlen et al. (2004), but at $z\approx2$ predict a significantly
higher value than they have reported at $z=1.6$.  Further
investigation into the rates at these high redshifts is required
before additional constraints may be placed on the relationship
between SFR and SNR, which is necessary to place limits on possible
progenitor pathways.  A confirmation of the plateau and decline of the
Type Ia SNR at $z>1$ would indicate a substantial ($t\approx3-4$ Gyr)
delay time.  Current measurements at $z\approx1$ and below are not
capable of conclusively determining the delay times, and are
consistent with a short ($t\lesssim1$ Gyr) time delay with respect to
the SFR.

\acknowledgments

We thank Tomas Dahlen and Brian Schmidt for helpful discussions
during the writing of this paper.  We further thank the anonymous 
referee for the many comments that greatly improved the manuscript.  
Financial support for this work was
provided by NASA through program GO-09118 from the Space Telescope
Science Institute, which is operated by the Association of
Universities for Research in Astronomy, Inc., under NASA contract NAS
5-26555.  Further support was provided by the National Science
Foundation through grant AST-0206329.

\clearpage

\begin{deluxetable}{cccccc}
\tablewidth{0pc}
\tablecaption{IfA Deep Survey Supernova Rates}
\tablehead{
\colhead{$z$\tablenotemark{a}} & \colhead{N\tablenotemark{b}} & \colhead{N\tablenotemark{c}} &
\colhead{N$_{predict}$\tablenotemark{d}} & \colhead{Rate\tablenotemark{e}} &
\colhead{$\sigma$(Rate)\tablenotemark{f}}}
\startdata
 0.25 &  9  &  1  &  5.9  &  0.17  & 0.17 \\ 
 0.35 &  9  &  5  &  9.5  &  0.53  & 0.24 \\ 
 0.45 & 15  &  9  & 12.3  &  0.73  & 0.24 \\ 
 0.55 & 40  & 29  & 14.2  &  2.04  & 0.38 \\ 
 0.65 & 25  & 23  & 15.4  &  1.49  & 0.31 \\ 
 0.75 & 32  & 28  & 15.7  &  1.78  & 0.34 \\ 
\enddata
\tablenotetext{a} {Center of redshift bin.  All redshift bins are of
width $z=0.1$.}
\tablenotetext{b} {Number of SNIa within redshift bin for sample
of 133 (before culling likely contaminants)}
\tablenotetext{c} {Number of SNIa within redshift bin for sample of 98}
\tablenotetext{d} {Predicted number of SNIa detectable by the IfA Deep
Survey within redshift bin for a constant rate of 10$^{-4}$ SNIa Mpc$^{-3}$ yr$^{-1}$ $h_{70}^3$}
\tablenotetext{e} {Type Ia SNR, calculated as N$^c$/N$_{predict}$, in units of 10$^{-4}$ SNIa Mpc$^{-3}$
yr$^{-1}$ $h_{70}^3$}
\tablenotetext{f} {SNR uncertainty, calculated as
(N$^c$)$^{0.5}$/N$_{predict}$}
\label{table:rates}
\end{deluxetable}

\begin{deluxetable}{cccccc}
\tablewidth{0pc}
\tablecaption{SFR/SNR Ratios}
\tablehead{
\colhead{$t$(Gyr)\tablenotemark{a}} & \colhead{$\chi^{2}/N$} &
\colhead{SFR/SNR($\sigma$)\tablenotemark{b}} &
\colhead{N(WD)($\sigma$)\tablenotemark{c}} & 
\colhead{SN/WD($\sigma$)\tablenotemark{d}}}
\startdata
 0.0  &    3.7  &   450(200)  &   9.5(4.2)  & 0.11(0.05)   \\
 0.5  &    3.4  &   520(240)  &  10.9(5.0)  & 0.09(0.04)   \\
 1.0  &    2.2  &   580(280)  &  12.2(5.9)  & 0.08(0.04)   \\
 1.5  &    2.8  &   690(310)  &  14.5(6.5)  & 0.07(0.03)   \\
 2.0  &    5.5  &   905(380)  &  19.0(8.0) & 0.05(0.02)   \\
 2.5  &    5.4  &   970(590)  &  20.4(12.4) & 0.05(0.03)   \\
 3.0  &    6.6  &  1060(770)  &  22.3(16.2) & 0.04(0.03)   \\
 3.5  &    8.4  &  1170(820)  &  24.6(17.2) & 0.04(0.03)   \\
 4.0  &   10.3  &  1380(910)  &  29.0(19.1) & 0.03(0.02)   \\
 4.5  &    9.6  &  1430(1000) &  30.0(21.0) & 0.03(0.02)   \\
\enddata
\tablenotetext{a} {Linear time delay (Gyr) applied to SFR values}
\tablenotetext{b} {SFR/SNR in units of M$_{\mathord\odot}$/SNIa $h_{70}^{-2}$ }
\tablenotetext{c} {Number of C-O white dwarfs (with uncertainty) produced by the amount
of star formation indicated by column $b$ (see Section ~\ref{cowd})}
\tablenotetext{d} {Fraction of C-O white dwarfs (with uncertainty) that result in Type Ia
SN, assuming a single-degenerate pathway}
\label{table:sfr22}
\end{deluxetable}

\clearpage

\clearpage

\epsscale{1.0}

\clearpage

\clearpage

\epsscale{1.0}

\clearpage

\clearpage

\clearpage

\clearpage

\clearpage

\end{document}